\documentclass[floatfix,twocolumn,showpacs,preprintnumbers, amsmath,amssymb,prb]{revtex4}
\usepackage{subfigure}
\usepackage{graphicx}
\usepackage{dcolumn}
\usepackage{bm}
\usepackage[latin1]{inputenc}
\begin{document}
\preprint{APS}
\bibliographystyle{aip}

\title{
Magnetism of Substitutional Co Impurities in Graphene: Realization of 
Single $\pi$-Vacancies}

\author{E.~J.~G.~Santos}
\email{eltonjose_gomes@ehu.es}
\affiliation{
Centro de F\'{\i}sica de Materiales,
Centro Mixto CSIC-UPV/EHU, Apdo. 1072,
20080 San Sebasti\'an, Spain}
\affiliation{Donostia International Physics Center (DIPC),
Paseo Manuel de Lardizabal 4, 20018 San Sebasti\'an, Spain}

\author{D. S\'anchez-Portal}
\email{sqbsapod@ehu.es}
\affiliation{
Centro de F\'{\i}sica de Materiales,
Centro Mixto CSIC-UPV/EHU, Apdo. 1072,
20080 San Sebasti\'an, Spain}
\affiliation{Donostia International Physics Center (DIPC),
Paseo Manuel de Lardizabal 4, 20018 San Sebasti\'an, Spain}

\author{A. Ayuela}
\email{swxayfea@ehu.es}
\affiliation{
Centro de F\'{\i}sica de Materiales,
Centro Mixto CSIC-UPV/EHU, Apdo. 1072,
20080 San Sebasti\'an, Spain}
\affiliation{Donostia International Physics Center (DIPC),
Paseo Manuel de Lardizabal 4, 20018 San Sebasti\'an, Spain}

\date{\today}

\begin{abstract}
We report  {\it ab initio} calculations of  the structural, electronic
and magnetic properties of a graphene monolayer substitutionally doped
with Co (Co$_{sub}$)  atoms. We focus in Co  because among traditional
ferromagnetic elements  (Fe, Co and Ni), only  Co$_{sub}$ atoms induce
spin-polarization in graphene. Our results show that, nearby the Fermi
energy, the electronic structure of substitutional Co is equivalent to
that   of    a   vacancy    in   a   $\pi$-tight-binding    model   of
graphene. Additionally,  we investigate the  magnetic coupling between
several Co$_{sub}$.  We  find that the total spin  moment of arrays of
Co$_{sub}$ defects  depends on the Co  occupation of A  and B graphene
sublattices,  in  accord  with  the  Lieb'  s  theorem  for  bipartite
lattices. The magnetic couplings show more complex behaviors.
\end{abstract}


\maketitle

The peculiar electronic and magnetic properties of graphene monolayers
have               recently               attracted               much
attention.~\cite{Novoselov04,Novoselov05,Geim07,graph-review}       New
types  of electronic  and, particularly,  spintronic devices  based on
graphene  have been  proposed.   These graphene  structures have  also
driven  an increasing  interest  to study  defects,  which are  always
expected  in real-life  devices.  Intrisic  defects have  been already
widely
studied~\cite{graph-review,palacios06,amara07,lehtinen03,palacios08}
and extrinsic defects, like  substitutional atoms, are presently under
intense research.   \cite{peres07,Pereira08, Wehling07, 1st} Recently,
Gan {\it  et al.}~\cite{Banhart08} using  high-resolution transmission
electron  microscopy (HRTEM)  have observed  substitutional Au  and Pt
atoms in graphene monolayers.  They observed large activation energies
for   in-plane  migration,   which  point   towards   strong  covalent
carbon-metal  bonding  and  high  stability  for  such  substitutional
defects.  In spite of the  growing interest about defects in graphene,
little is known on magnetic properties, such as spin polarizations and
couplings,  for the  traditional bulk  magnets  Fe, Co  and Ni  placed
substitutionally in graphene.

In this  work, we  present a study  of the structural,  electronic and
magnetic  properties  of substitutional  Co  (Co$_{sub}$)  atoms in  a
graphene monolayer.  We show that Co has spin polarization. This is in
contrast to  Fe and Ni  substitutions which surprinsingly do  not show
spin  polarization  \cite{elsewhere}.  We  observe  that  there  is  a
one-to-one  correspondence between  the expected  behavior  for single
vacancies in a  simple $\pi$-tight binding model of  graphene and that
found for the Co$_{sub}$ defects.   The Co atom stabilizes a symmetric
structure  of the  carbon  vacancy. The  electronic  structure of  the
Co$_{sub}$ defect at the Fermi  energy (E$_{Fermi}$) is dominated by a
single level with a strong contribution from the $p_z$ orbitals of the
neighboring C  atoms.  Each Co$_{sub}$  defect shows a spin  moment of
1.0 $\mu_B$.  The total  spin, however, follows closely Lieb's theorem
for bipartite  lattices~\cite{lieb89} and depends on the  number of Co
substitutions  in  each  sublattice.   Thus,  magnetic  couplings  are
predominantly (anti)ferromagnetic for  Co$_{sub}$ defects sited in the
(opposite) same sublattice.  Our calculations also show the dependence
of the couplings on the crystalline direction and relative position of
the defects.

\begin{figure}
\includegraphics[width=2.2400in]{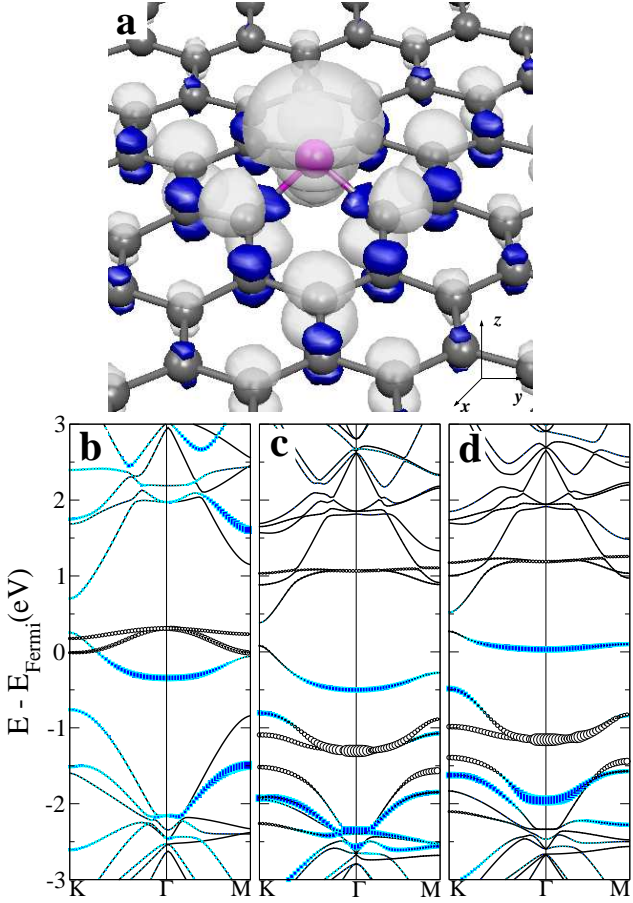}
\includegraphics[width=3.27300in]{fig1bcd.eps}
\caption{(Color online) (a) Isosurface  of the spin density induced by
a Co$_{sub}$  defect. Positive and negative  spin densities correspond
to   light   and   dark   surfaces  with   isovalues   of   $\pm$0.008
e$^-$/Bohr$^3$, respectively.  Panel (b) presents the spin-unpolarized
band structure of an unreconstructed carbon vacancy. Panel (c) and (d)
show, respectively,  the majority and  minority spin band  structure a
Co$_{sub}$  defect in  a similar  cell.  Filled  symbols in  panel (b)
indicate  the contribution  of the  $p_{z}$  orbitals of  the C  atoms
surrounding  the  vacancy, whereas  empty  symbols  correspond to  the
$sp^{2}$ character.  In  panels (c) and (d), filled  and empty circles
denote  the contribution  of  hybridized Co~$3d_{z^2}$-C~$2p_{z}$  and
Co~$3d$-C~$2sp^{2}$ characters, respectively. E$_F$ is set to zero.}
\label{fig1}
\end{figure}

We   have  used  the   SIESTA  code~\cite{portal97,soler02}   for  our
calculations.     We   have    checked   that    the   norm-conserving
pseudopotential~\cite{troullier91}   used    for   Co   satisfactorily
reproduces  the  band structure  and  spin  moment  of bulk  Co.   The
pseudopotential includes nonlinear core corrections~\cite{louie82} for
exchange and  correlation, with a pseudo-core matching  radius of 0.67
a.u.  Computational  parameters guarantee the  convergence of relative
energies   for   different    magnetic   configurations   within   the
meV.~\cite{1st}    The    results     obtained    with    VASP    code
~\cite{kresse93,kresse96}, using a well converged plane-wave cutoff of
400~eV and  the projected-augmented-wave scheme,  are almost identical
to those obtained with SIESTA.

The relaxed  geometry and the  magnetization density for  a Co$_{sub}$
defect in a graphene monolayer is given in Fig.~\ref{fig1}(a).  The Co
atom appears displaced 0.92~\AA\ from the carbon surface, and it has a
symmetrical  three-fold coordinated structure  with Co-C  distances of
1.77~\AA. The  three C-Co-C bond angles  are quite similar  and have a
value $\sim$97$^o$.  The  binding energy of the Co  atom to the carbon
vacancy  in  graphene is  7.6  eV.   In  comparison, the  most  stable
adsorption configuration of Co on  graphene has a much lower energy of
1.3~eV.   Therefore, Co$_{sub}$  defects  are likely  to  be found  in
carbon  nanostructures synthesized  using Co-containing  catalysts, as
seems to  be the case  for Ni~\cite{ushiro06}.  Furthermore  one could
expect that Co$_{sub}$  defects can be fabricated in  a controlled way
by creating vacancies in  graphene and subsequently depositing Co. The
presence  of Co$_{sub}$ defects  could be  detected today  by scanning
tunneling   microscopy    (STM)~\cite{stm-paper},   X-ray   adsorption
techniques~\cite{ushiro06} and HRTEM~\cite{Banhart08}.

Figure~\ref{fig1}~(a) shows  the spin-polarization pattern  induced by
the presence  of a Co$_{sub}$ impurity. The  spin polarization induced
in the  neighboring carbon  atoms has a  clear $p_z$-like  shape.  The
sign  of the  spin  polarization follows  the  bipartite character  of
graphene: the polarization aligns  parallel (antiparallel) to the spin
moment of  the Co  impurity for carbon  atoms sitting in  the opposite
(same)  sublattice  than  the  Co  atom.  The  total  spin  moment  is
1.0~$\mu_{B}$. Using Mulliken  population analysis, 0.44~$\mu_{B}$ are
assigned to  the Co  atom; 0.20~$\mu_{B}$, to  the three  first carbon
neighbors; and  0.36~$\mu_{B}$, delocalized in the rest  of the layer.
The spin polarization decays with distance very slowly.

\begin{figure}
\includegraphics[width=2.800in]{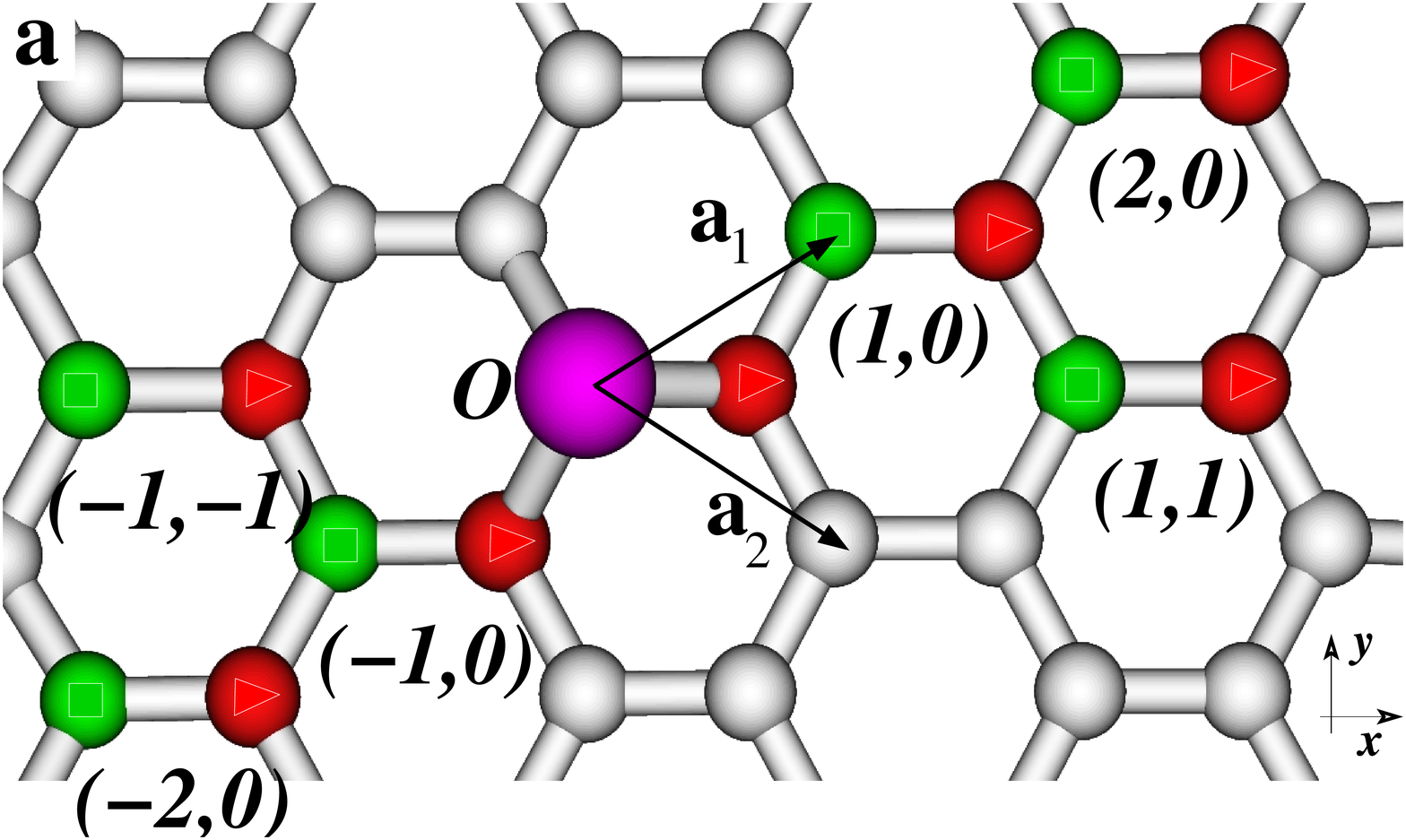}
\includegraphics[width=2.800in]{fig2b.eps}
\includegraphics[width=2.800in]{fig2c.eps}
\caption{(Color online)  (a) Schematic representation  of the geometry
used  to  calculate  the  relative stability  of  ferromagnetic  (FM),
antiferromagnetic  (AFM)  and spin  compensated  (PM)  solutions as  a
function of the positions of two Co$_{sub}$ impurities.  Sublattices A
and B  are indicated by  squares and triangles, respectively.   One of
the impurities is fixed in a central A-type site, whereas the other is
moved along  the (b)  $(n,n)$ and (c)  $(n,0)$ directions.   The empty
circles represent spin  compensated solutions and the full circles
correspond to a fit with a Heisenberg model (see text).}
\label{fig2}
\end{figure}

To understand the origin of  this spin polarization, we now analyze in
detail the band structure  of this system.  Figures~\ref{fig1} (c) and
(d)  present the  results  for  a Co$_{sub}$  defect  in a  4$\times$4
graphene supercell.  Similar results are obtained  using an 8$\times$8
cell. For  comparison, the panel  (b) shows the  spin-compensated band
structure of  a single unreconstructed  carbon vacancy, the  so called
$D_{3h}$  vacancy.~\cite{amara07}  For  the  $D_{3h}$  vacancy,  three
defect  states appear  in a  range  of $\sim$0.7~eV  around the  Fermi
energy  (E$_F$).   Two  states  appearing  at 0.3~eV  above  E$_F$  at
$\Gamma$ have  a large contribution from  the $sp^{2}$ lobes  of the C
atoms surrounding the vacancy. Other state at 0.35 eV below E$_{F}$ at
$\Gamma$ shows  a predominant  $p_z$ contribution.  The  $p_{z}$ level
corresponds with the defect state that appears pinned at E$_{F}$ for a
vacancy in a simplified $\pi$-tight-binding description of graphene.

When a  Co atom is bound  to the vacancy, the  defect states described
above hybridize with the $3d$  states of Co.  The two $2sp^{2}$ defect
bands,  now an  antibonding combination  of Co~$3d$  and  the original
C~$2sp^{2}$   vacancy   levels,  are   pushed   at  higher   energies,
$\sim$1.0~eV  above  E$_{F}$ (see  Fig.~\ref{fig1}~(c)  and (d)).  The
singly  occupied $p_{z}$  state,  now hybridized  mainly  with the  Co
$3d_{z^2}$ orbital,  remains pinned at the E$_{F}$  and becomes almost
fully  spin-polarized.    Thus,  the  Co$_{sub}$   impurities  becomes
analogous  to a  $\pi$ vacancy.   The splitting  between  majority and
minority $p_{z}$ defect  bands is $\sim$ 0.50 eV.   The spin splitting
is much smaller for all other bands.

We consider next the  magnetic couplings between Co$_{sub}$ defects in
a  large  8$\times$8  supercell  with two  Co$_{sub}$  impurities.  We
calculate the energy difference between different spin alignments as a
function of  the relative position of  the defects.  Figure~\ref{fig2}
shows  the  results  along  with  a schematic  representation  of  our
notation.  These energies are  comparable to the spin-flip energies in
ferromagnetic transition  metals Fe, Co and  Ni.  Several observations
can be  made: ({\it i})  when the impurities  are located in  the same
sublattice (AA  systems) the ferromagnetic (FM)  configuration is more
stable  than the  antiferromagnetic (AFM)  one; ({\it  ii}) if  the Co
atoms are in opposite sublattices (AB systems) it is very difficult to
reach  a FM solution,~\cite{AB-FM-Comments}  instead the  system finds
either  a spin-compensated  (PM) or  an AFM  solution; ({\it  iii}) at
short  distances ($<$  3.0~\AA) the  systems always  converge  to spin
compensated solutions.

\begin{figure}
\includegraphics [width=2.6500in]{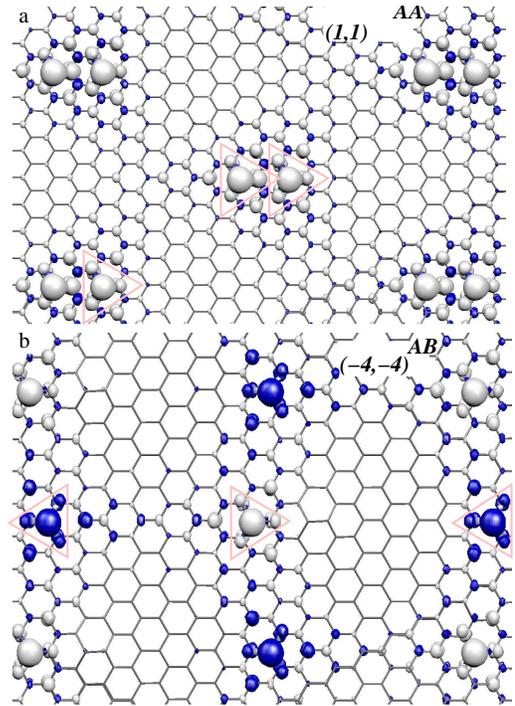}
\caption{  (Color online)  (a) Spin  densities for  configurations (a)
$(1,1)^{AA}$,  and (b)  $(-4,-4)^{AB}$(see Fig.~\ref{fig2}(a)  for the
nomenclature).  Positive and negative  spin densities are indicated by
light (gray)  and dark (blue) isosurfaces  corresponding to $\pm$0.001
e$^-$/Bohr$^3$, respectively.  }
\label{fig3}
\end{figure}

 In  the FM  cases, the  total spin  magnetization integrates  to 2.00
 $\mu_{B}$,  with a  spin population  on  every Co  atom that  remains
 almost  constant $\sim$0.50 $\mu_{B}$  ($\sim$0.30 $\mu_{B}$  for the
 three C-nearest neighbors).   In other cases the total  spin is zero.
 Thus, the  total magnetic moment  of the system follows  the equation
 $S=|N_{sub}^{A}-N_{sub}^{B}|$,  where N$_{sub}^{A(B)}$ is  the number
 of Co$_{sub}$ defects in  the A(B) sublattices.  Therefore, our total
 moment   is   consistent  with   Lieb'   s   theorem  for   bipartite
 lattices~\cite{lieb89}.  This   supports  our  analogy   between  the
 electronic structure of Co$_{sub}$  defects and single vacancies in a
 simplified $\pi$-tight-binding description of graphene.

The  spin magnetization  density for  some selected  configurations is
plotted in Fig.~\ref{fig3}.  Although  the spin is fairly localized on
the Co  atom and the neighboring  C atoms, the presence  of the defect
also causes a delocalized  magnetization density with alternated signs
in  the two sublattices.   The triangular  pattern, that  reflects the
three-fold symmetry  of the layer, shows different  orientations for A
and B substitutions. This explains the very anisotropic AB interaction
along the  $(n,n)$ direction seen  in Fig.~\ref{fig2} (b):  the energy
difference   between   AFM   and   PM   solutions   for   $(n,n)^{AB}$
configurations  strongly  depends  on  the relative  position  of  the
impurities,  showing  such   directionality.   Similar  patterns  have
already                          been                         observed
experimentally~\cite{Halas98,Foster89,Rutter07,Ruffieux00}   in  point
defects (e.g.  vacancies and  hydrogens) on graphene  by means  of STM
techniques      and     also      theoretically      discussed     for
$\pi$-vacancies.~\cite{Oleg-Yazyev08,Palacios08,Pereira08}          For
Co$_{sub}$  in graphene,  similar STM  experiments should  display the
topology of the spin densities given in Fig.~\ref{fig3}.

We  have  also  investigated  the  magnetic  interactions  within  the
framework   of   a   classical   Heisenberg  model,   $H=   \sum_{i<j}
J_{AA/AB}({\bf r}_{ij}){\bf  S}_{i}{\bf S}_{j} $,  where ${\bf S}_{i}$
is  the local  moment  for a  Co$_{sub}$  impurity at  site $i$.   The
expression for the ${\bf r}_{ij}$  dependence of the exchange has been
taken     from      analytical     RKKY     coupling      given     in
Ref.~\onlinecite{Saremi07},  except for the  exponent of  the distance
decay, which is  fitted to our {\it ab  initio} results.  The exchange
interaction  for AA systems  can be  fitted with  a $|r_{ij}|^{-2.43}$
distance dependence  (see the full  circles in Fig.\ref{fig2}  (b) and
(c)).  This distance dependence  is in  reasonable agreement  with the
$|r_{ij}|^{-3}$   behavior  obtained   with   analytical  models   for
substitutional defects  and voids~\cite{Saremi07,Voz05}.  In  the case
of  AB systems a  simple RKKY-like  treatment fails  to satisfactorily
describe the interactions, at least for the relatively short distances
between defects considered in our calculations.

Next, we  explain the appearance  of PM solutions  in Fig.~\ref{fig2}.
The  appreciable  interaction  between  defect levels  in  neighboring
impurities for AB systems opens a {\it bonding-antibonding} gap in the
$p_z$  defect band~\cite{gap-comment}  and, thus,  contributes  to the
stabilization of PM solutions.  For AA systems, however, the bipartite
character  of  the  graphene  lattice makes  the  interaction  between
defects  much  smaller.  This  explains the  prevalence  of  solutions
showing  a local  spin polarization  for AA  configurations.  For very
short distance between  impurities, a larger defect-defect interaction
opens a  large gap and, in consequence,  stabilizes PM configurations.
It is interesting to point  out that similar behaviors were previously
observed  for vacancies  in  graphene and  graphene ribbons  described
within            a           simplified           $\pi$-tight-binding
model.~\cite{Oleg-Yazyev08,Kumazaki07,Palacios08}

Similar behaviors to Co$_{sub}$ are expected for any other system that
can be  mapped into  a simple $\pi$-vacancy  in graphene. This  is the
case   of  a   hydrogen  atom   saturating  a   C  $p_{z}$   state  in
graphene.~\cite{katsnelson-h,oleg-h-vacancy}  However,  there are  two
important  differences  between  these  two systems.   As  Co  prefers
energetically to sit on  carbon vacancies, Co$_{sub}$ defects could be
fabricated by depositing  Co atoms on areas where  vacancies have been
previously  created by, for  example, electron  irradiation or  Ar ion
bombardment.  The calculated diffusion barrier of hydrogen on graphene
is 1.19~eV,~\cite{h-diffusion}  the experimentally observed activation
energy for the migration of  substitutional metals, such as Au and Pt,
in graphene  is around 2.5~eV.~\cite{Banhart08}.   Thus, the diffusion
is also probably much slower  for Co$_{sub}$ defects than for hydrogen
atoms.  We expect that stable arrays of Co$_{sub}$ impurities could be
fabricated  and  open  a  route  to  obtain in  reality  many  of  the
interesting magnetic behaviors  obtained for $\pi$-models of defective
graphene.

In summary, only Co substitutional  atoms in graphene respect to other
traditional ferromagnetic  materials, such as Fe and  Ni, present spin
polarization.  The  electronic structure  of  Co$_{sub}$ defects  (see
Fig.~\ref{fig1})  is dominated by  an impurity  level at  E$_{F}$ with
strong contribution from the $p_z$  states of the neighboring C atoms,
and it  is a  clear reminiscent  of a single  vacancy in  a simplified
$\pi$-tight-binding  description of  graphene.   The study  of the  Co
coupling in several sites demonstrates FM alignments for the AA cases,
in  contradistinction  to  AB  cases  which are  either  AFM  or  spin
compensated. The  AA cases  were fitted to  a Heisenberg model  with a
RKKY exchange  constant. Our  findings for Co$_{sub}$  can be  used to
bring into contact the  engineering of nanostructures with the results
of $\pi$-models in defective graphene.

We acknowledge support from Basque Dep. de Educaci\'on and the UPV/EHU
(Grant    No.    IT-366-07)     and    the    Spanish    MCI    (Grant
No. FIS2007-66711-C02-02).


\begin{thebibliography}{100}
\bibitem{gap-comment}
For the  AB systems,  we always find  bonding-antibonding gaps  in the
defect band ranging from 0.3 eV to 0.9~eV when going from $(1,1)^{AB}$
to the  $(-1,-1)^{AB}$ configuration, respectively.  These values have
to be compared with  the $\sim$0.5~eV spin-splitting of the Co$_{sub}$
defect.  All those  AB systems with gaps larger  than 0.4~eV are found
to converge to PM solutions.
\end{thebibliography}
\end{document}